\documentclass[12pt]{article}
\usepackage{epsfig}
\begin{document}
\begin{titlepage}
\begin{center}

\hfill{FTUV/05-0624}

\hfill{IFIC/05-27}

\hfill{ SNUTP/05-012}

\hfill{ APCTP/05-008}

\vspace{2cm}

{\Large \bf A QCD sum rule analysis of the pentaquark} \\
\vspace{0.50cm}
\renewcommand{\thefootnote}{\fnsymbol{footnote}}
Hee-Jung Lee $^{a,b}$\footnote{Heejung.Lee@uv.es},
N.I. Kochelev$^{c,d}$\footnote{kochelev@theor.jinr.ru},
V. Vento$^a$\footnote{Vicente.Vento@uv.es}
\vspace{0.50cm}

{(a) \it Departament de F\'{\i}sica Te\`orica and Institut de F\'{\i}sica Corpuscular,\\
Universitat de Val\`encia-CSIC, E-46100 Burjassot (Valencia),
Spain}  \vskip 1ex
 {(b) \it Asia Pacific Center for Theoretical Physics,
POSTECH, Pohang 790-784, Korea} \vskip 1ex {(c) \it School of
Physics and Center for Theoretical Physics, \\Seoul National
University, Seoul 151-747, Korea}
 \vskip 1ex
{(d) \it Bogoliubov Laboratory of Theoretical Physics,\\
Joint Institute for Nuclear Research, Dubna, Moscow region, 141980
Russia}

\end{center}
\vskip 0.5cm

\centerline{\bf Abstract} We perform a QCD sum rule calculation to
determine the mass and the parity of the lowest lying pentaquark
state. We include operators up to dimension $d=13$ in the OPE and
the direct instanton contributions. We find evidence for a
positive parity state. The contribution from operators of
dimension $d>5$ is instrumental in determining the parity of the
state and achieving the convergence of the sum rule. \vskip 0.3cm
\leftline{Pacs: 12.38.Aw, 12.38.Lg, 12.39.Ba, 12.39.-x}
\leftline{Keywords: quarks, instanton, hadrons, pentaquark}

\vspace{1cm}
\end{titlepage}

\setcounter{footnote}{0}

\section{Introduction}
 The experimental and theoretical status of $\Theta^+$-pentaquark
remains controversial \cite{exp}, \cite{jaffe}. The QCD sum rules
(SRs) have shown to be a very powerful tool for the investigation
of the properties of conventional \cite{QCDSR} and exotic
multiquark hadronic states \cite{pivovarov}. Several attempts to
describe the properties of $\Theta^+$ pentaquark using SRs have
appeared \cite{zhu,sugiyama,eidemuller,narison,matheus,ioffe}.
However, these calculations were either restricted to low
dimension operators \cite{zhu,sugiyama,eidemuller,narison} or they
used interpolating currents which did not have the most suitable
quantum numbers to project onto the $\Theta^+$
\cite{matheus,ioffe}.

The dynamics associated with the instanton, the 't Hooft
interaction, has been successful in understanding the spectroscopy
of four-quark \cite{multidor} and H-dibaryon \cite{kochdorH,okaH}
states and it was important for the spectroscopy of the pentaquark
\cite{klv}. Moreover, instantons are crucial for understanding
chiral symmetry breaking in the strong interactions \cite{shuryak,diakonov}
and lie at the basis of chiral soliton model for
baryons, which has predicted the pentaquarks and their peculiar
properties, e.g. small widths and masses \cite{DPP}.

In the SR calculations thus far, the contribution from so-called
direct instantons, has not been investigated
\footnote{The direct
instanton contribution of ref.~\cite{dirwrong} should vanish due to
the Pauli principle for the quarks in the instanton field.}. It is
well known that direct instantons play an important role in the
SRs calculations to determine the properties of the pseudoscalar
mesons and the nucleon octet baryons \cite{shuryak1,dorkoch1}. We
showed, in a mixed model-SR calculation, that they might be also
important for the pentaquarks \cite{lkv}.

We perform a calculation for the pentaquark SRs which takes into
account operators up to dimension $d=13$ and direct instanton
contributions, and leads to evidence for a positive parity state
whose mass is close to the observed mass.

\section{ The standard OPE contribution to the sum rules}
The QCD sum rule approach starts from the correlator of some
relevant current,
\begin{equation}
\Pi(q^2)=i\int d^4x \ e^{iq\cdot x}\langle 0| T
\eta_{\Theta}(x)\bar{\eta}_{\Theta}(0)|0\rangle=\hat{q}\Pi_1(q^2)+\Pi_2(q^2)\ .
\label{corr}
\end{equation}
Here $\eta_{\Theta}$ represents a current with non vanishing
projection onto the pentaquark state. We use the conventional
notation $\hat{q}=\gamma\cdot q$.

The use of the narrow resonance approximation,
\begin{equation}
{\rm Im}\Pi(q^2)=\pi\lambda^2_\Theta(\hat{q}+M_\Theta)\delta(q^2-M_\Theta^2)
+\theta(q^2-s_0^2)[\ \hat{q}{\rm Im}\Pi_1(q^2)+{\rm Im}\Pi_2(q^2)\ ]\ ,
\end{equation}
where $M_\Theta$ is the mass of the pentaquark, $\lambda_\Theta$
its residue, $s_0$ the threshold, and the appropriate dispersion
relations lead to the so-called chirality even
\begin{equation}
\frac{1}{\pi}\int_0^{s_0^2} ds^2 \ e^{-s^2/M^2}
{\rm Im}\Pi_1^{OPE}(s^2) =\lambda_\Theta^2e^{-M_\Theta^2/M^2}\ ,
\label{even}
\end{equation}
and chirality odd
\begin{equation}
\frac{1}{\pi}\int_0^{s_0^2} ds^2 \ e^{-s^2/M^2}
{\rm Im}\Pi_2^{OPE}(s^2) =\lambda_\Theta^2M_\Theta e^{-M_\Theta^2/M^2}\
\label{odd}
\end{equation}
sum rules.

Our choice of current in the pentaquark correlator is
\begin{equation}
\eta^{A}_{\Theta}=\frac{1}{4\sqrt{2}}\
\epsilon_{afg}\epsilon_{abc}\epsilon_{bde}
[(u^T_d C d_e)\gamma_5C\bar{s}_c^T][u^T_{f} C\gamma_5 d_{g}] \ ,
\label{currentA}
\end{equation}
whose structure corresponds to the A--state of
refs.~\cite{klv,lkv}, which consists of the product of a scalar
$ud\bar{s}$ triquark and a pseudoscalar $ud$--diquark. It can be
easily seen, that this current has the same structure as that of
ref.~\cite{sugiyama} except for the $\gamma_5$ in front of the
strange quark field. The consequence of this similarity is that
for the chirality odd sum rule our results become identical to
their results, if we restrict the calculation to low dimension
operators, take into account our different normalization and an
additional negative sign due to negative intrinsic parity of our
current, Eq.~(\ref{currentA}).

We have also considered the current,
\begin{equation}
\eta^B_{\Theta}=\frac{1}{4\sqrt{6}}\ \epsilon_{acd} [(u_a^TC\gamma_\mu
d_b+u_b^TC\gamma_\mu d_a) \gamma_5\gamma^\mu C\bar{s}_b^T]
[u_c^TC\gamma_5d_d]\ ,
\end{equation}
which corresponds to the B--state of refs.~\cite{klv,lkv} and
contains a vector $ud\bar{s}$ triquark and a scalar $ud$--diquark.
This current has also negative intrinsic parity. However, our
analysis has shown that the B current coupling with $\Theta^+$ is
weak and therefore no definitive conclusion about the values of
the mass and residue can be drawn from the consideration of its
correlator. We will therefore not discuss it further.

Let us proceed with the analysis of the chirality odd SR for the
pentaquark, Eq.~(\ref{odd}). This SR is directly related to the
mass of the state and usually is more stable than the chirality
even SR for the ground state baryons \cite{dorkoch1} and triquark
$ud\bar s$ states \cite{lkv}.

After Borel transforming the pentaquark correlator has dimension
$d=13$. The calculation of the chirality odd sum rule will be
performed taking into account operators up to dimension $d=13$. To
this order we obtain good stability in the OPE result. The
operators of higher dimension ($d>13$)
appear in the sum rule multiplied by inverse powers of the
Borel mass, thus, in the interesting Borel mass region, their
contribution is small and can be safely neglected.

\begin{figure}[t]
\centerline{\epsfig{file=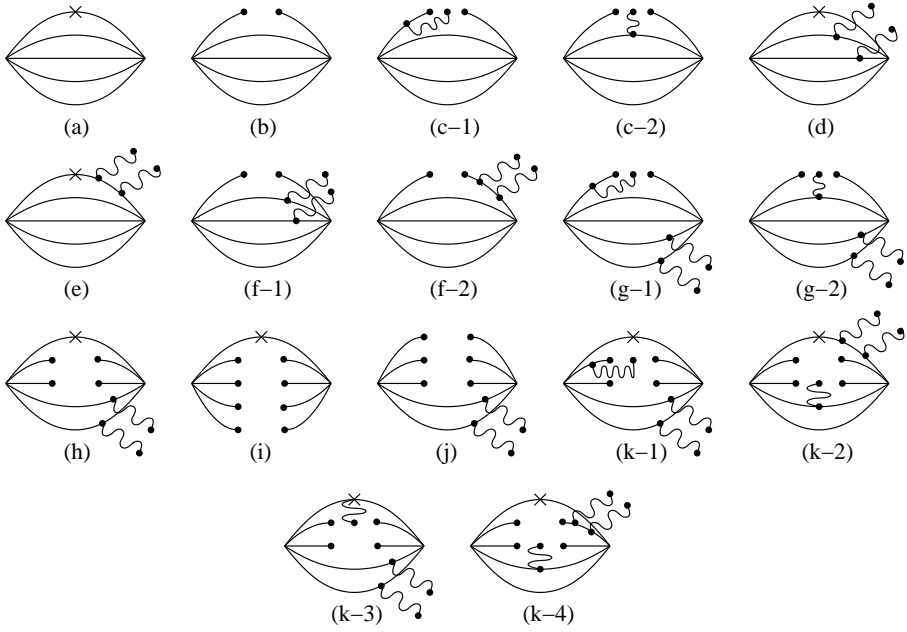,width=12cm,angle=0}}\
\caption{Diagrams contributing to the chirality odd pentaquark sum
rule in our calculation. The quark propagator on top corresponds
to the strange quark and $\times$ denotes a quark mass insertion.}
\label{dA}
\end{figure}

\begin{figure}
\centerline{\epsfig{file=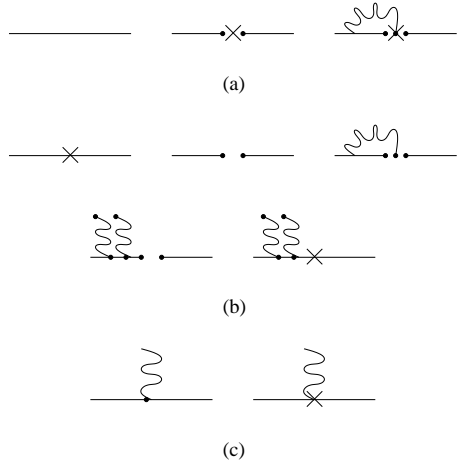,width=6cm,angle=0}}\
\caption{(a) The terms $F_1^q$ in the quark propagator, (b) the
terms $F_2^q$ in the propagator, (c) two last terms in
Eq.~\ref{prop}.}
\label{propagator}
\end{figure}

With our interpolating current, the relevant trace to the
chirality odd sum rule is expressed as
\begin{eqnarray}
 &&{\rm Tr}\langle 0|
T\eta_\Theta(x)\bar{\eta}_\Theta(0)|0\rangle_{odd}
=-(i)^5\frac{1}{32}\epsilon_{abc}\epsilon_{bde}\epsilon_{afg}
\epsilon_{a'b'c'}\epsilon_{b'd'e'}\epsilon_{a'f'g'}
{\rm Tr}[\gamma_5CS^{s,T}_{c'c}(-x)C\gamma_5]
\nonumber\\
&&\times\bigg({\rm Tr}[CS^{u,T}_{dd'}CS^d_{ee'}] {\rm
Tr}[CS^{d,T}_{gg'}C\gamma_5S^u_{ff'}\gamma_5] +{\rm
Tr}[CS^{u,T}_{df'}CS^d_{eg'}\gamma_5] {\rm
Tr}[CS^{d,T}_{ge'}C\gamma_5S^u_{fd'}]
\nonumber\\
&&-{\rm Tr}[CS^{u,T}_{df'}CS^d_{ee'}CS^{u,T}_{fd'}C
\gamma_5S^d_{gg'}\gamma_5] -{\rm
Tr}[CS^{u,T}_{dd'}CS^d_{eg'}\gamma_5CS^{u,T}_{ff'}C
\gamma_5S^d_{ge'}]\bigg)
\end{eqnarray}
where the superscripts on the quark propagator mean the quark
flavor and $a,b,c,...,$ are the color indices. In Fig.~\ref{dA}
the diagrams which contribute to the chirality odd SR up to $d=13$
are shown. In order to calculate the correlator to a certain order
we need to consider the quark propagator to the appropriate
dimension. In Fig.~\ref{propagator} we show the corresponding OPE
diagrams for the quark propagator which lead to
\begin{eqnarray}
S^q_{ab}(x)&=&-i\langle 0|Tq_a(x)\bar{q}_b(0)|0\rangle
\nonumber\\
&=&\delta_{ab}(\hat{x}F_1^q+F_2^q)-i\tilde{g}G^{\mu\nu}_{ab}\frac{1}{x^2}
(\hat{x}\sigma_{\mu\nu}+\sigma_{\mu\nu}\hat{x})
\nonumber\\
&&-m_q\tilde{g}G^{\mu\nu}_{ab}\sigma_{\mu\nu}\bigg(\ln(-x^2\Lambda^2/4)+2\gamma_{EM}\bigg),
\label{prop}
\end{eqnarray}
$a,b$ are the color indices and ${\tilde g}=g_c/32\pi^2$. The two
functions are given by
\begin{eqnarray}
F_1^q&=&\frac{1}{2\pi^2 x^4}+\frac{m_q \langle
\bar{q}q\rangle}{48}
+i\frac{m_qx^2}{2^7\cdot3^2}g_c\langle\bar{q}\sigma\cdot G
q\rangle,\nonumber\\
F_2^q&=&i\frac{m_q}{4\pi^2x^2}+i\frac{\langle \bar{q}q\rangle}{12}
-\frac{x^2}{192}g_c\langle\bar{q}\sigma\cdot G q\rangle
+i\frac{g_c^2x^4}{2^{9}\cdot3^3}\langle \bar{q}q\rangle\langle
G^2\rangle
\nonumber\\
&&+i\frac{m_qg_c^2}{2^9\cdot 3\pi^2}\langle G^2\rangle x^2
\bigg(\ln(-x^2\Lambda^2/4)+2\gamma_{EM}-\frac{2}{3}\bigg)\ ,
\end{eqnarray}
where $\gamma_{EM}$ is the Euler--Mascheroni constant and we take
 $\Lambda=500$ MeV~\cite{pasupathy87}. Note, that for massless
$u$, $d$ quarks, $F_i^u=F_i^d$.

Our result for the SR including  operators up to dimension $d=13$
has the form
\begin{eqnarray}
&&\frac{1}{4}\bigg[-\frac{1}{15}m_sM^{12}E_5\bigg|_{(a)}
-\frac{2}{15}f_saM^{10}E_4\bigg|_{(b)}
+\frac{1}{6}f_sm_0^2aM^8E_3\bigg|_{(c)}
\nonumber\\
&&-\frac{1}{12}bm_sM^8E_3\bigg|_{(d)}
-\frac{1}{12}bm_sM^8W_3\bigg|_{(e)}
-\frac{4}{27}bf_saM^6E_2\bigg|_{(f)}
\nonumber\\
&&+\frac{1}{12}f_sm_0^2abM^4E_1\bigg|_{(g)}
-\frac{1}{12}m_sba^2M^2E_0\bigg|_{(h)}
+\frac{8}{27}m_sa^4\bigg|_{(i)}
\nonumber\\
&&-\frac{1}{72}bf_sa^3\bigg|_{(j)}\
+\frac{1}{48}m_sm_0^2ba^2\bigg|_{(k)}\bigg]
=\tilde{\lambda}^{2}_\Theta M_\Theta e^{-M_\Theta^2/M^2}\ .
\label{SRF}
\end{eqnarray}

Each term corresponds to a diagram in Fig.~\ref{dA}. The
residue is defined by
$\tilde{\lambda}_\Theta=(4\pi)^4\lambda_\Theta$. The contributions
from the continuum are given by the following functions :
\begin{eqnarray}
E_n(M)&=&\frac{1}{\Gamma(n+1)M^{2n+2}}\int_0^{s_0^2}dx\
e^{-x/M^2}x^n\ ,
\nonumber\\
W_n(M)&=&\frac{1}{\Gamma(n+1)M^{2n+2}}\int_0^{s_0^2}dx\
e^{-x/M^2}x^n \bigg(-2\ln(x/\Lambda^2)+\ln\pi
\nonumber\\
&&+\psi(n+1)+\psi(n+2)+2\gamma_{EM}-\frac{2}{3}\bigg)
\end{eqnarray}
with $\psi(n)=1+1/2+\cdots+1/(n-1)-\gamma_{EM}.$ The numerical
values of various quantities in the sum rule will be given in
below.

We have checked that our result Eq.~(\ref{SRF}) taking into
account only  contributions of operators with dimensions $d\leq 9$
is in agreement with the previous calculations \cite{sugiyama},
\cite{narison}. We should mention that due to specific spin and
color structure of our current Eq.~(\ref{currentA}) the
potentially important contributions from operators
$\langle\bar{\psi}\psi\rangle^3$,
$\langle\bar{\psi}\psi\rangle^2\langle\bar{\psi}g_c\sigma\cdot
G\psi\rangle$,  $m_s\langle g_c\bar{q}\sigma\cdot G q\rangle^2 $,
and $\ \langle\bar{s}s\rangle\langle g_c\bar{q}\sigma\cdot G
q\rangle^2$ can not appear in chirality odd sum rules. The absence
of the contribution of these operators can be seen by direct
analysis of the terms in Eq. (7). We would like to mention that
their appearance depends strongly on the structures of the
interpolating current. For example, in the recent paper
\cite{oganesian},  where another pentaquark current has been used,
it was shown that these operators give non-vanishing contribution.
We do not include the contributions proportional to $\langle
g_c^3GGG\rangle$ and $\langle g_c^2GG\rangle^2 $, which in the OPE
are related to terms of higher orders in the expansion in the
strong coupling constant, and therefore their contributions are
expected to be very small for the light quark systems (see for
example the discussion in ref. \cite{narison} on the three-gluon
condensate contribution). This statement is in the agreement with
a general observation that pure gluonic operators are not very
important in QCD SRs for hadrons consisting of light $u-, d-,$ and
$s-$ quarks \cite{rly}, \cite{narison2}.

\section{The direct instanton contribution to the sum rule}
In addition to contributions of power type, arising from the OPE
expansion, there are {\it exponential} contributions coming from
direct instantons contributions to the
correlators~\cite{shuryak1,dorkoch1}. They can be calculated by
using the following formula for the quark propagator in the
instanton background in the regular gauge
\begin{equation}
S^{q,inst}_{ab}(x,y)=A_q(x,y)\gamma_\mu\gamma_\nu (1+\gamma_5)
(U\tau^-_\mu\tau^+_\nu U^\dagger)_{ab},
\end{equation}
where $$ A_q(x,y)=-i\frac{\rho^2}{16\pi^2
m_q^{*}}\phi(x-z_0)\phi(y-z_0) $$
and
$$\phi(x-z_0)=\frac{1}{[(x-z_0)^2+\rho^2]^{3/2}}.$$
Note that $\rho$ stands for the instanton size
and $z_0 $ the center of the instanton ; $U$ represents the color
orientation matrix of the instanton in $SU(3)_c$ and
$\tau_{\mu,\nu}$ are $SU(2)_c$ matrices ;
$m_q^*=m_{cur}^q-2\pi^2\rho_c^2\langle\bar q q\rangle/3$ is the
effective quark mass in the instanton vacuum and $m_{cur}^q$ the
current quark mass. The final result should be multiplied by a
factor of two to take into account the anti-instanton
contribution, and has to be integrated over the instanton density.

To leading order in the instanton density, the direct instanton
contributions arise from two body $ud$, $u\bar s$, $d\bar s$ and
three body $ud\bar s$ quark zero mode propagators in the
correlator Eq.(\ref{corr}), as shown in Fig. 3.

\begin{figure}[htb]
\centering
\vspace*{0.5cm}
\psfig{file=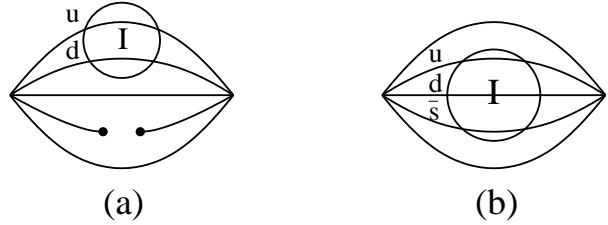,width=8cm,height=3cm,angle=0}
\caption{An example of instanton two- and three-body contributions
to the correlator of the pentaquark currents.}
\end{figure}

The final result for two body instanton contribution is
\begin{eqnarray}
\Pi_2(M)&=& -\frac{n_{eff}\rho_c^4\langle \bar{q}
q\rangle}{2^6\cdot3\pi^{8}m_q^*m_s^*} \hat{B}[f_6(Q)],
\label{qs}
\end{eqnarray}
where Shuryak's instanton liquid model for QCD vacuum with density
$n(\rho)=n_{eff}\delta(\rho-\rho_c)$ \cite{shuryak} has been used
and $ \hat{B}[f_6(Q)] $ is the Borel transform of $f_6(Q)$ which
is defined by
\begin{equation}
f_6(Q)=\int d^4z_0\int d^4x \frac{e^{-iq\cdot
x}}{x^6[z_0^2+\rho_c^2]^3[(x-z_0)^2+\rho_c^2]^3}\ ,
\label{f6}
\end{equation}
where $\rho_c$ is the average instanton size. There are two types
of singularities in Eq.(\ref{f6}). One of them is related to the
pole at the origin $x^2=0$, the other is due to the pole at finite
distance from origin $x^2\sim -\rho_c^2$. The pole at $x^2=0$
produces, after Fourier transforming, power terms in $1/Q^n$ in
addition to the exponential type direct instanton contributions
$\exp(-Q\rho_c)$, arising from finite distances. One should
carefully subtract that contribution to avoid double counting with
the standard OPE terms. We follow the procedure suggested in ref.
\cite{shifman} for the analysis of direct instanton contributions
to heavy quark decay. More specifically, for a general integral
\begin{equation}
\Pi_{ins}= \int
d^4xd^4z_0e^{iqx}\frac{S(x)}{x^{2n}((x-z_0)^2+\rho_c^2)^{\alpha}(z_0^2+\rho_c^2)^{\beta}},
\end{equation}
where $S(x)$ contains no singularities for complex $x_\mu$, we use
Feynman's parameterization
\begin{equation}
\Pi_{ins}=\frac{\Gamma(\alpha+\beta)}{\Gamma(\alpha)\Gamma(\beta)}
\int
d^4xd^4z_1e^{iqx}\frac{S(x)}{x^{2n}}\int_0^1dt\frac{t^{\alpha-1}(1-t)^{\beta-1}}
{(t(1-t)x^2+z_1^2+\rho_c^2)^{\alpha+\beta}}, \label{integral}
\end{equation}
where $z_1=z_0-tx$, but consider only the contribution from
the pole at
\begin{equation}
x^2=-(z_1^2+\rho_c^2)[t(1-t)]^{-1}.
\end{equation}
The Borel transform of the function $f_6$ is given by
\begin{eqnarray}
\hat{B}[f_6(Q)]&=&-\frac{\pi^4M^{12}}{2^{13}}\int_0^1dt\int_0^1
dy\ \frac{e^{-M^2\rho_c^2/(4ty(1-y))}}{y^2(1-y)^2}
\bigg(X^2+5X^3+10X^4
\nonumber\\
&&+10X^5+5X^6+X^7\bigg)\ ,
\end{eqnarray}
where $X=(1-t)/t$. Note that only the contribution from the pole
at finite quark separation has been considered.

We have also performed the calculation of the three body
contributions induced by instantons, Fig. 3b, and found that they
vanish as the result of the cancellation between diagrams with
different $ud\bar s$ combinations in the pentaquark current. The
two body  $ud$ instanton contribution also cancels. The only non
vanishing contributions arise from the two body $u\bar s$ and $d\bar s$
terms of Eq.~(\ref{qs}). The behaviour of the different
contributions is associated to the Dirac structure of our current
Eq.~(\ref{currentA}) which includes both scalar and pseudoscalar
$ud$ diquarks with equal weights. The instanton induced
contribution is very sensitive to the parity of the state
\cite{carlitz} and it flips sign when the parity of the state
changes. Of course, the three-body instanton induced forces might
give non-zero contribution for other choices of pentaquark
currents, for example, for currents with derivatives.

We should mention that three body instanton
terms induce forces which give non zero contributions to the mass
of some specific triquark $ud\bar s$ pentaquark
clusters~\cite{klv,lkv} and furthermore, they give non vanishing
contribution to the $\Theta$ mass within the bag
model~\cite{bagOka}.

\section{Numerical analysis}
We use the following values for parameters at the normalization
point 2 GeV~\cite{narison} (see also recent discussion about
uncertainties  in values of  various condensates in~\cite{ioffe2})

\begin{eqnarray}
&&\langle\bar u u\rangle=-(243\ {\rm MeV})^3\equiv-\frac{a}{(2\pi)^2}, \nonumber\\
&& b=\langle g_c^2G^2\rangle=0.88\ {\rm GeV}^4,\nonumber\\
&&ig_c\langle\bar{u}\sigma\cdot Gu\rangle=m_0^2\langle \bar{u} u\rangle
=0.8{\rm GeV}^2\langle\bar{u} u\rangle,
\nonumber\\
&&\frac{\langle \bar{s} s\rangle}{\langle \bar{u} u\rangle}=
\frac{\langle\bar{s}\sigma\cdot G
s\rangle}{\langle\bar{u}\sigma\cdot G u\rangle}=f_s=0.8 ,\nonumber
\\
&& m_s=111\ {\rm MeV},  \label{par}
\end{eqnarray}
with $\rho_c=1.6$ GeV$^{-1}$ for the average instanton size in
the QCD vacuum. In the numerical estimate of the direct instanton
contribution the relations of the instanton liquid model
\cite{DEK}
\begin{equation}
\frac{n_{eff}}{m_q^{*2}}=\frac{3}{4\pi^2\rho_c^2}, \ \ \
\frac{m_q^*}{m_s^*}=\frac{1}{f_s-\frac{3m_s}{2\pi^2\rho_c^2\langle\bar{q}q\rangle}}\ ,
\end{equation}
are used.

\begin{figure}
\begin{minipage}[c]{7cm}
\vspace*{1.2cm}
\hspace*{0cm}\psfig{file=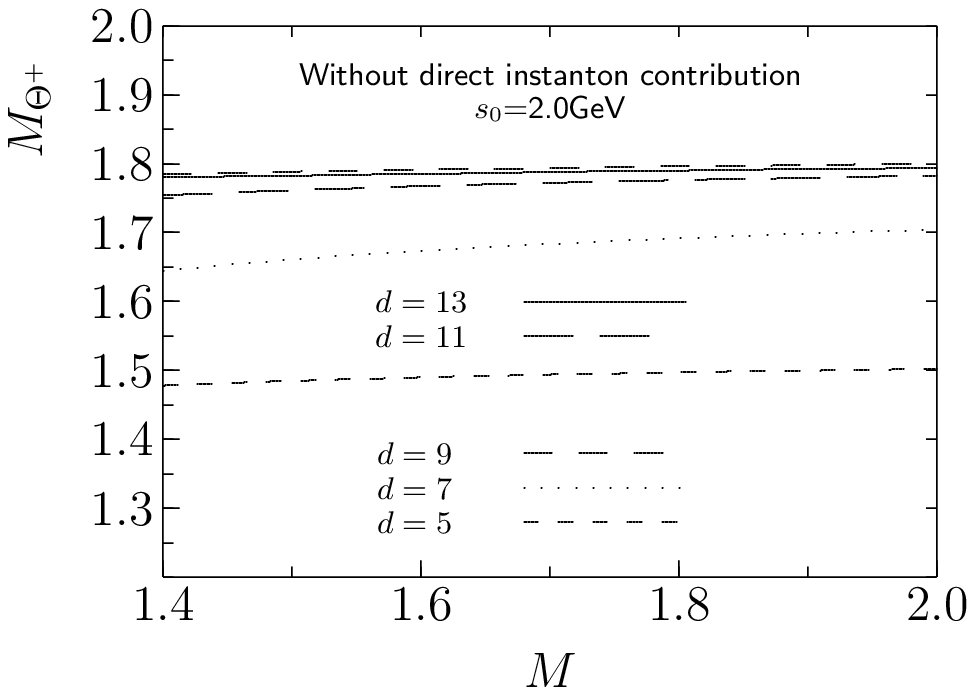,width=6cm,height=6cm}
\caption{The mass of the pentaquark  obtained without direct
instanton contributions as a function of the Borel parameter in
different orders of the OPE expansion.}
\end{minipage}
\hspace*{0.5cm}
\begin{minipage}[c]{7cm}
\hspace*{0.5cm} \vskip 1.2cm
\psfig{file=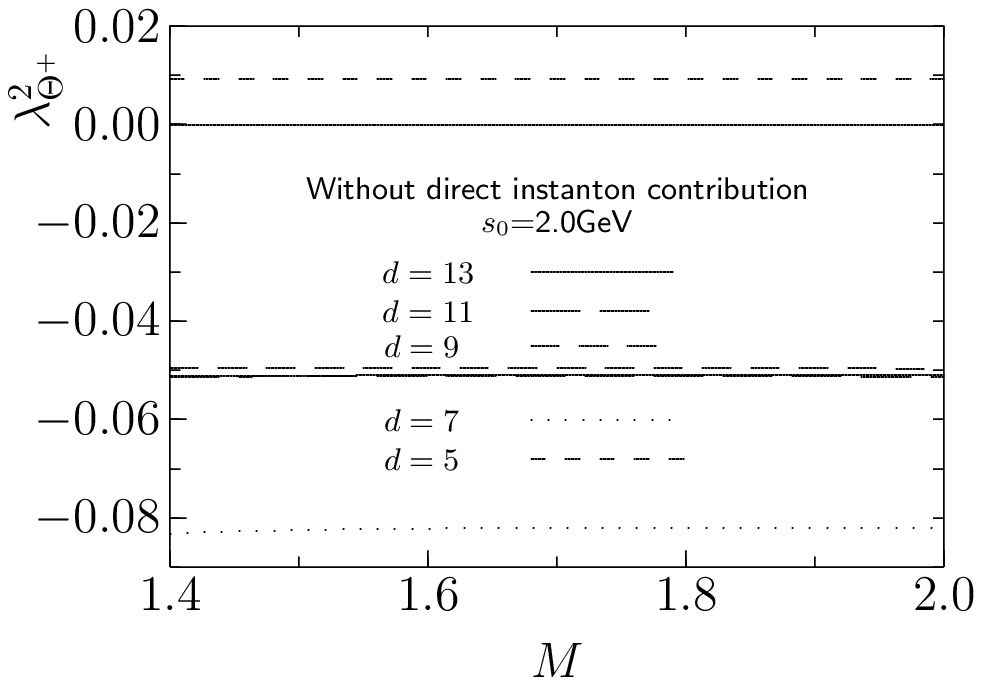,width=6cm,height=6cm} \caption{The residue
of the pentaquark obtained without direct instanton contribution
as a function of the Borel parameter in different orders of OPE
expansion.}
\end{minipage}
\end{figure}

In Figs.~4 and 5 the result of the calculation of the pentaquark
mass and residue within the standard OPE expansion for the
different orders in operator dimensions is shown. In Figs.~6 and 7
the mass and residue of the pentaquark as a function of the value
of the Borel parameter with direct instanton contributions are
shown. In Fig.~8 we present the results of the calculation of the
OPE and the direct instanton contributions to the left-hand side
of the SR, Eq.(\ref{odd}). All curves are given for a value of the
threshold $s_0=2$ GeV. We chose this value of threshold because
the stability was best. From the fit of the sum rules we arrive at
the following values for pentaquark mass:
$M_{\Theta^+}=1.66$ GeV for $d=7$, $M_{\Theta^+}=1.75$GeV for
$d=9$, $M_{\Theta^+}=1.73$GeV for $d=11$, and $M_{\Theta^+}=1.75$
GeV for $d=13$~\footnote{Once the instanton contribution is
included the stability in the Borel parameter for the SR up to
$d=5$ operators disappears (see Fig. 6). Therefore it is not
possible to extract the value of pentaquark mass from the SR with
only up to $d=5$ operators.}.

One important result of our calculation is in the change of the
sign of the squared of the residue when increasing the dimension
of the operators which contribute to the OPE. Thus, for $d=5$ the
sign is positive, while it becomes negative for higher dimensions.
In particular, the contribution from the dimension $d=7$ operators
is crucial for inverting the sign. Due to the negative intrinsic
parity of our current Eq.~(\ref{currentA}), the negative
(positive) sign of the squared of the residue implies positive
(negative) parity for the state. Therefore, our final result for
the residue presented in Fig.~7 shows that one can arrive to the
wrong conclusion about the parity of the pentaquark state~\cite{sugiyama},
if one takes the decision based on only the
contributions of low dimension ($d=5$) operators. We also stress
the necessity to include high dimension operators to get good
convergence for the sum rule. It is evident that this effect is
directly related to the high dimension of the pentaquark current.

Once we include the contributions from the high dimension
operators and the instantons, our result for the $\Theta^+$
pentaquark mass, $M_{\Theta^+}\approx 1.75$ GeV, is  higher than
was given by previous SR
calculations~\cite{sugiyama,zhu,eidemuller} but still in rough
agreement with the available experimental data, if one admits
about 10\% accuracy in the predictions of the SR approach due to
uncertainties in the values of the various  condensates, the mass
of the strange quark, the contribution from higher dimension
operators $d > 13$, higher order pQCD corrections,  etc.
Furthermore, some additional effects such as the  mixing between
various pentaquark states ~\cite{klv,lkv}, which are beyond the
scope of the present paper, might give some additional
contribution to the mass of the $\Theta^+$.

\begin{figure}
\begin{minipage}[c]{8cm}
\hspace*{0.5cm} \psfig{file=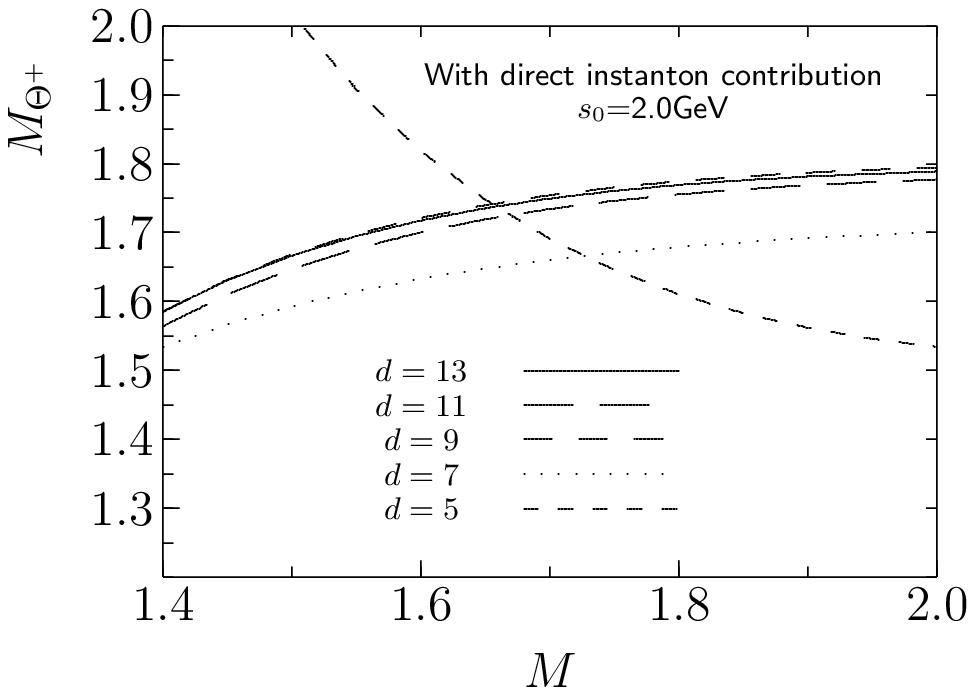,width=6cm,height=6cm}
\caption{The mass of the pentaquark obtained with direct instanton
contributions as a function of the Borel parameter for different
orders of OPE expansion.}
\end{minipage}
\hspace*{0.5cm}
\begin{minipage}[c]{8cm}
\hspace*{0.5cm} \psfig{file=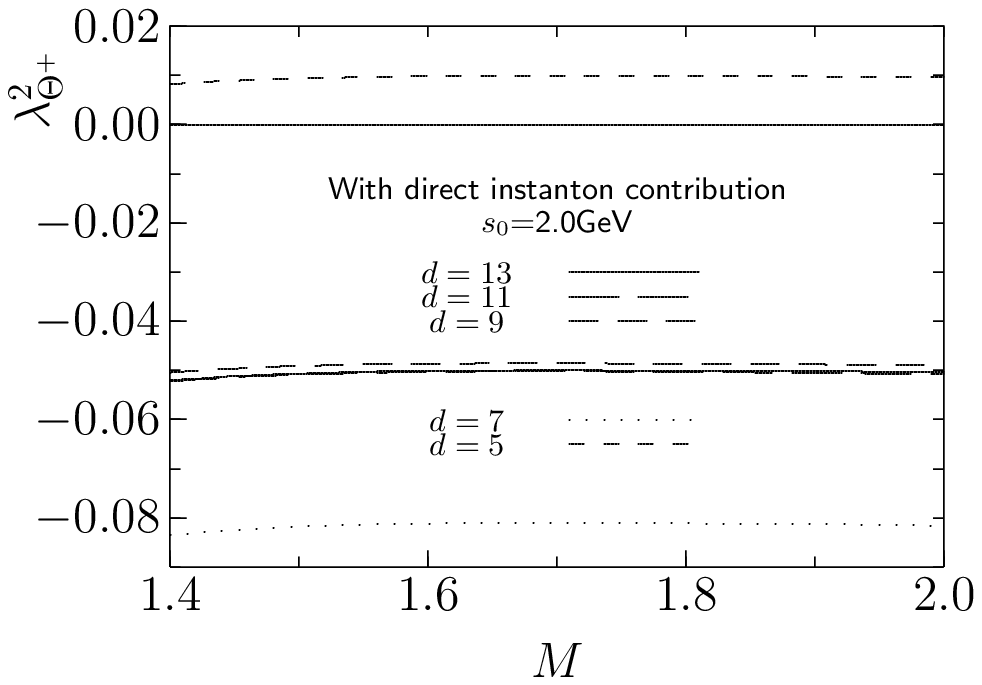,width=6cm,height=6cm}
\caption{The residue of the pentaquark obtained with direct
instanton contribution as a function of the Borel parameter in
different orders of the OPE expansion. }
\end{minipage}
\end{figure}

\begin{figure}
\centering
\hspace*{0cm} \psfig{file=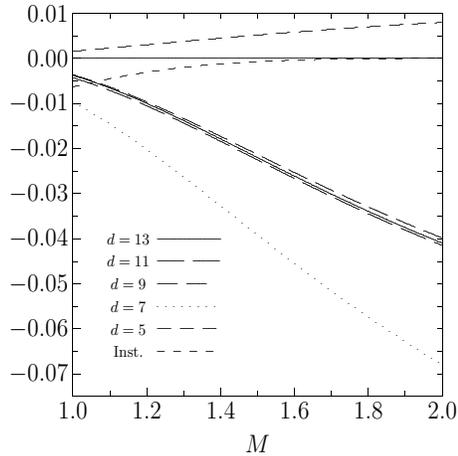,width=6cm,height=6cm}
\caption{The different OPE orders and  direct instanton
contributions to the left-hand side of chirality odd SR . }
\end{figure}
We also note that in our calculation the pentaquark has positive
parity in agreement with the soliton model prediction \cite{DPP}.
Our estimate for direct instanton contribution is done within
Shuryak's instanton liquid model. We have found that the instanton
contribution for the full SR is rather small, but can give a large
contribution to it when one considers operators only up to
dimension $d=5$  (see Figs.~4,6). The smallness of the instanton
contribution to the full SR is mainly related to the large mass of
the Borel parameter $M\approx 1.7$ GeV, where we obtain the
plateau of stability (Fig. 6). In this region the instanton
contribution is small in comparison with the contribution from the
high dimension operators in the OPE (Fig. 8).

There is  a significant dependence of our results on the value of
threshold. This is a common feature in all the studies about the
properties of the pentaquark within the QCD sum rule approach. In
our case,  we have chosen  $s_0=2$ GeV  to satisfy the physical
requirement of having a large stability plateau.

In summary, we have shown the analysis of the QCD sum rules for
the $\Theta^+$ pentaquark current including high dimension
operators in the OPE and direct instanton contributions. Our
results conclude that the role of the high dimension operators is
important for obtaining a positive parity for pentaquark state.
Our calculation though produces a bound state whose mass is higher
than the experimental observation. More sophisticated models and
probably states mixing~\cite{klv,lkv} might reduce the obtained
value to the observed one.

\section{Acknowledgments}
We are grateful to A.E. Dorokhov, S.V. Esaibegian, S.V. Mikhailov,
A.G. Oganesian, and A.A. Pivovarov for very useful discussions.
This work was supported by grants
 by MCyT-FIS2004-05616-C02-01 and
GV-GRUPOS03/094~(VV), by
 Ministerio de Educacion y Ciencia of
Spain (SEEU-SB2002-0009) (HJL), and by Brain Pool program of Korea
Research Foundation through KOFST, grant 042T-1-1 and in part by
RFBR-03-02-17291, RFBR-04-02-16445 (NIK). NIK is very grateful to
the School of Physics, SNU for their warm hospitality during the
final stage of this work.

\end{document}